\documentclass[sigconf]{acmart}

\usepackage{booktabs} 

\usepackage{lmodern}
\usepackage{amsfonts}
\usepackage{amssymb}
\usepackage{xspace}
\usepackage{epstopdf}
\usepackage{enumitem}
\usepackage{enumerate}
\usepackage{algorithm}
\usepackage{algorithmic}
\usepackage{multirow}
\usepackage{balance}
\usepackage{booktabs}
\usepackage{array}
\usepackage{wrapfig}
\usepackage{pifont}
\usepackage{pbox}
\usepackage{geometry}
\usepackage{color}
\usepackage{balance}
\newcommand{\ts}{\textsuperscript}

\renewcommand\footnotetextcopyrightpermission[1]{} 
\newcommand\blfootnote[1]{%
  \begingroup
  \renewcommand\thefootnote{}\footnote{#1}%
  \addtocounter{footnote}{-1}%
  \endgroup
}

\begin{document}

\fancyhead{}

\def\x{{\mathbf x}}
\def\L{{\cal L}}
\def\x{{\mathbf x}}
\def\L{{\cal L}}
\def\eg{\textit{e.g.}}
\def\ie{\textit{i.e.}}
\def\Eg{\textit{E.g.}}
\def\etal{\textit{et al.}}
\def\etc{\textit{etc.}}
\def\nback{\textit{n}-back }

\title{EEG-based Evaluation of Cognitive Workload Induced by Acoustic Parameters for Data Sonification}

\author{Maneesh Bilalpur}
\affiliation{%
 \institution{CVIT, IIIT Hyderabad}
 \country{India}
}
\email{maneesh.bilalpur@research.iiit.ac.in}

\author{Mohan Kankanhalli} 
\affiliation{%
 \institution{School of Computing, National University of Singapore}
 \country{Singapore}}
\email{mohan@comp.nus.edu.sg}

\author{Stefan Winkler}
\affiliation{%
  \institution{Advanced Digital Sciences Center, University of Illinois at Urbana-Champaign}
  \city{Singapore} 
  \country{Singapore}}
\email{stefan.winkler@adsc-create.edu.sg}

\author{Ramanathan Subramanian}
\affiliation{%
  \institution{Advanced Digital Sciences Center, University of Illinois at Urbana-Champaign}
  \city{Singapore} 
    \country{Singapore}
}
\email{ramanathan.subramanian@ieee.org}


\renewcommand{\shortauthors}{M. Bilalpur, M. Kankanhalli, S. Winkler, and R. Subramanian}
\renewcommand{\shorttitle}{EEG-based Evaluation of Cognitive Workload Induced by Acoustic Parameters}

\begin{abstract}
Data Visualization has been receiving growing attention recently, with ubiquitous smart devices designed to render information in a variety of ways. However, while evaluations of visual tools for their interpretability and intuitiveness have been commonplace, not much research has been devoted to other forms of data rendering, \eg, sonification.
This work is the first to automatically estimate the cognitive load induced by different acoustic parameters considered for sonification in prior studies~\cite{ferguson2017evaluation,ferguson2018investigating}.  We examine cognitive load via (a) perceptual data-sound mapping accuracies of users for the different acoustic parameters, (b) cognitive workload impressions \textit{explicitly} reported by users, and (c) their \textit{implicit} EEG responses compiled during the mapping task. Our main findings are that (i) low cognitive load-inducing (\ie, more intuitive) acoustic parameters correspond to higher mapping accuracies, (ii) EEG spectral power analysis reveals higher $\alpha$ band power for low cognitive load parameters, implying a congruent relationship between explicit and implicit user responses, and (iii) Cognitive load classification with EEG features achieves a peak F1-score of 0.64, confirming that reliable workload estimation is achievable with user EEG data compiled using wearable sensors.
\blfootnote{Accepted for publication in the proceedings of the 20\ts{th} ACM International Conference on Multimodal Interaction, Colorado, USA.\\DOI:10.1145/3242969.3243016}
\end{abstract}


\maketitle

\section{Introduction}
Understanding and sensemaking from multi-dimensional data is a challenge, since the traditional medium for visual data representation and communication is typically restricted to two or three dimensions. This calls for visualization and content delivery tools utilizing alternative sensing mechanisms such as auditory~\cite{kendall1991visualization}, tactile~\cite{lee2007tactile}, gustatory and olfactory~\cite{ranasinghe2011digital}. Also, given the significant proportion of visually challenged persons the world over (around 253 million people are visually impaired\footnote{http://www.who.int/en/news-room/fact-sheets/detail/blindness-and-visual-impairment}), employing purely visual communication techniques makes information inaccessible to a large section of the society.

Nevertheless, attempts to use non-visual modalities for encoding data attributes are few. Among these, data sonification, where data attributes are conveyed via psychoacoustic signals, is a relatively mature technique with applications in multiple fields such as astrophysics and neurology~\cite{ferguson2018investigating}. Two recent works that investigate the suitability/ease-of-understanding of data-to-sound mappings are~\cite{ferguson2017evaluation} and~\cite{ferguson2018investigating}. In~\cite{ferguson2017evaluation}, a study evaluating auditory parameters (such as pitch, roughness, noise and sharpness) that best convey the focus level of an astronomical image is presented. This study is further extended in~\cite{ferguson2018investigating}, where the perceptual congruence between three data attributes, namely, \textit{stress}, \textit{error} and \textit{danger} and the aforementioned acoustic parameters is evaluated. Both evaluations are based on explicit user assessments acquired via the mouse and keyboard.

A limiting factor in real-life situations where visualizations are put to use is that explicit user feedback may not be available for improving or adapting the data rendering methodology. Acquiring user feedback via \textit{implicit} means would therefore be critical for optimal information communication. Neuroergonomics, which examines human factors by employing neuroscientific methods, presents a viable alternative in this regard. This approach is also attractive with the advent of light-weight, wearable sensors. A number of works have explored \textit{cognitive sensing} of users presented with visual information-- the user's level of (dis)comfort with the presented information is gauged via eye movements~\cite{huang2007using,raschke2014visual} or neural activity in the form of EEG~\cite{anderson2011user,wang2016using} and fNIRS~\cite{peck2013using}. Such studies have however not been performed, to our knowledge, for auditory perception tasks.

This paper builds on prior works~\cite{ferguson2017evaluation,ferguson2018investigating}, and seeks to estimate the cognitive load of users mapping acoustic parameters to data attributes via implicit means. Cognitive load induced by different auditory parameters is examined via explicit as well as implicit user responses-- participants completed (a subset of questions from) the NASA-TLX questionnaire~\cite{hart1988development} to convey their impressions regarding how \textit{easy} (low cognitive load condition) or \textit{difficult} (high cognitive load condition) it is to map the focus level of an astronomical image to an acoustic parameter; their EEG responses while performing the mapping task were also recorded via the commercial \textit{Emotiv} device. The main findings of this work are the following: (1) Perceptual (focus level to acoustic data) mapping accuracies are higher for parameters inducing low cognitive load as per explicit user impressions; (2) On performing power spectral analysis for the different EEG spectral bands, higher $\alpha$ band power is noted for low cognitive load inducing parameters, implying a congruence between explicit and implicit user responses; (3) On segregating acoustic parameters as \textit{low} or \textit{high} cognitive load inducing based on user impressions, we attempted  cognitive load classification for each auditory presentation trial by examining the EEG data. Experiments confirm that a maximum F1-score of 0.64 is achievable with a CNN based classifier. This implies that user perception of acoustic data can be gauged via implicit means. 
Overall, this work makes the following contributions. 
\begin{itemize}
\item[i.] To our knowledge, this is the first work to expressly investigate the cognitive load induced by multiple acoustic parameters via explicit and implicit means. Our analysis combines examination of the (a) perceptual mapping accuracies for data sonification, (b) user cognitive load impressions, and (c) user EEG responses.
\item[ii.] We demonstrate congruence between explicit user impressions and implicit neural activity, Acoustic parameters inducing low cognitive load are found to be associated with higher EEG $\alpha$ power, in line with previous findings~\cite{wang2016using}.
\item[iii.] We show that better-than-chance cognitive load classification is possible by examining the user EEG signals. A maximum F1-score of 0.64 is obtained for low-vs-high cognitive load classification using a deep convolutional neural network (CNN) classifier.
\end{itemize}

The rest of the paper is organized as follows. Section~\ref{RW} motivates our work in the context of available literature, and highlights its novelty. Section~\ref{ED} details the experimental design and protocol. Section~\ref{ERA} examines explicit user data in terms of perceptual mapping accuracies and cognitive load impressions, while Section~\ref{EEG} presents an analysis of the EEG data. In Section~\ref{CLC} we introduce the cognitive load classification experiments and discuss the results, while Section~\ref{DC} concludes the paper.

\section{related work}\label{RW}

This section reviews literature on data sonification and cognitive load estimation to motivate the need for our study.

\subsection{Visualization and Data Sonification}
Information Visualization (InfoViz) concerns the design and development of interactive and graphical representations of information. Interactive visualizations typically convey visual and data patterns utilizing rendering techniques that best cater to human perception and cognition~\cite{2009-protovis,Kranthi18}. Among the various visualization techniques, sonification involves the use of non-speech based audio signals to convey information. A sonification system conveys data values by manipulating acoustic parameters such as sound frequency (pitch) or tempo. Two recent works that attempt to identify the optimal acoustic parameters for conveying different (types of) data attributes are~\cite{ferguson2017evaluation} and~\cite{ferguson2018investigating}.

A study presented in  \cite{ferguson2017evaluation} explores the utility of \textit{sharpness}, \textit{roughness}, \textit{noise}, \textit{pitch} and a combination of \textit{roughness} and \textit{noise} for conveying the focus level of an astronomical image. The sound parameters are carefully chosen, upon reviewing sonification literature in great detail.
Sound parameters in \cite{ferguson2017evaluation} are evaluated by: (1) computing the mean perceptual data:sound mapping accuracy for focus level determination, and (2) comparing the performance of the sound parameters against the visual stimuli to evaluate if the acoustic parameters can indeed serve as effective proxies for conveying the visual information. The study concludes with two main findings: (a) Acoustic parameters that converge on a clear/pure sound are optimal for focus determination, and (b) The investigated auditory parameters can provide effective substitution for visual information.

An extension of the above study is presented in~\cite{ferguson2018investigating}. Here, the ability of parameters such as \textit{roughness}, \textit{noise} and \textit{pitch} for conveying negative data attributes such as \textit{error}, \textit{danger} and \textit{stress} is explored. This study concludes that the effectiveness of sound parameters is governed by the ease with which users can perceive the data:sound mapping. An earlier study detailed in~\cite{walker2005mappings} suggests that intuitive data:sound mappings may not result in the best user performance in terms of accuracy or response times.

\subsection{Cognitive Workload Estimation}
Cognitive workload (or mental workload) has been traditionally employed as a standard for measuring task difficulty by many previous works \cite{anderson2011user,bashivan2015learning,wang2016using}. In particular, there is a large body of work correlating how neural activity captured via EEG~\cite{wang2016using,bashivan2015learning} and fNIRS signals~\cite{peck2013using} can enable cognitive workload assessment. The advantage of employing neural signals such as EEG and fNIRS is that they can be captured via light-weight and wearable commercial devices, as against other physiological modalities that require data to be recorded with bulky and specialized lab equipment.

The utility of EEG for measuring cognitive workload has been demonstrated by many previous works \cite{wang2016using,bashivan2015learning}. The observation that changes in the EEG $\theta$ and $\alpha$ band power are indicative of memory load is made in~\cite{wang2016using}. Specifically, low cognitive load is found to be associated with higher $\alpha$ band power. The cognitive load induced by visualizations such as bar and quartile plots is studied via EEG analysis in \cite{anderson2011user}. NASA-TLX parameters are employed to explicitly obtain cognitive load ratings from users, and the authors demonstrate  prefrontal cortex activity captured during task performance is relevant to the working memory consumption. 

Deep learning for cognitive load estimation is proposed in~\cite{bashivan2015learning}. The ability of convolutional neural networks (CNNs) to  preserve spatial, spectral and temporal structure of EEG 
is exploited in this work. Spectral band maps are first synthesized from EEG data corresponding to low and high mental workload tasks, and automated classification of low/high cognitive load from EEG data is then attempted.

\subsection{Analysis of Related Work}
Upon reviewing related literature, one can note that (1) The evaluation of auditory parameters for data rendering is still an active area of research. While the authors of~\cite{ferguson2017evaluation} and~\cite{ferguson2018investigating} base their findings on perceptual mapping accuracies obtained from user responses, a direct estimation of user cognitive load is not attempted in either of these works; (2) While mental workload estimation has been explored via cognitive sensing as users perform visual processing, to the best of our knowledge we have not encountered equivalent studies for sonification.

In this regard, we present the first work towards estimating cognitive load of users by analyzing their EEG signals. A salient aspect of our work is that these EEG signals are captured via a light-weight and wearable \textit{Emotiv} device. It is reasonable to expect that users will be willing to use such sensors to feedback their cognitive state as they perform real-life perception tasks. The next section details the stimuli and protocol employed for estimation of sonification-induced cognitive workload.

\section{stimuli and experimental design}\label{ED}

For the purpose of this study, we used the acoustic parameters employed in \cite{ferguson2017evaluation}, namely, \textit{noise}, \textit{pitch} (pure sinusoidal tones in a C-major scale as in Experiment 2 of~\cite{ferguson2017evaluation}), \textit{roughness} and \textit{combination} of roughness and noise. Furthermore, we used the original astronomical images with varying focus levels for comparison. In this section, we describe the stimuli used and the adopted experimental protocol.

\subsection{Stimuli}

\begin{figure*}[t]
\includegraphics[width=\linewidth]{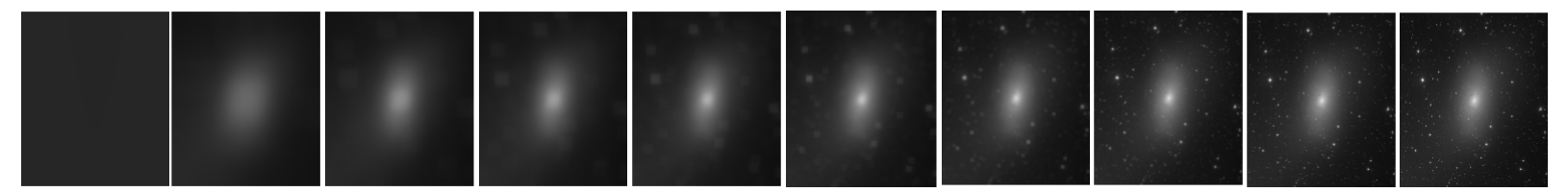}
\caption{\label{visual} Astronomical images shown with various focus levels. The image focus level progressively increases from 1--10 from left to right. \label{visual}}\vspace{-.2cm}

\end{figure*}

\subsubsection{Visual}
For benchmarking purposes, we used the astronomical images representing varying levels of focus in our study. These images are presented in Figure~\ref{visual}. A color image showing a telescopic view of the M110 galaxy was used for the highest focus level (level 10). Blurred versions of this image were generated using OpenCV via smoothing with different kernel sizes. The visual images are found to convey the focus level most precisely to users in \cite{ferguson2017evaluation}.

\subsubsection{Noise}
Similar to \cite{ferguson2017evaluation}, we used a 1000 Hz pure tone and broadband white noise to convey 10-levels of varying focus. As suggested in \cite{ferguson2017evaluation}, we exploit the possible association of noise with a negative attribute such as blur to denote the image focus level. A perfectly focused image (focus level of 10) corresponds to a 100\% pure tone while a 1-focus level (\ie, completely de-focused) image corresponds to 100\% noise.

\subsubsection{Pitch}
The sensation of acoustic frequency is referred to as Pitch. This parameter is usually found in music. Pure sinusoidal tones in a C-major scale (plus 2 extra notes to make the range 10 notes long) beginning at the middle C (C4, freq = 261.63 Hz) and ending on E6 (freq = 1318.51 Hz) were used. Loudness variations for the various frequencies were compensated for. Higher frequencies denoted higher focus levels.

\subsubsection{Roughness}
Having demonstrated in \cite{ferguson2017evaluation} the drawback of using a noisy carrier signal for roughness representation, we used a 100\% pure-tone that was amplitude modulated with 0, 2, 4, 7, 11, 16, 23, 34, 49 and 70 Hz. We believe that the dissonance in the acoustics best represents image blur. Hence we used a pure tone to represent the image of focus level-10.

\subsubsection{Combined Roughness and Noise}
This is the second best performing acoustic parameter in \cite{ferguson2017evaluation} after the visual images. Thus, we used this direct pairing of roughness and its corresponding noise. Analogous to the above described individual parameters, a pure-tone was used to represent the highest focus level of the image and the combination of 100\% noise modulated by 70Hz corresponds to focus level-1.

\subsubsection{Combined Image plus Acoustics}
In addition to the above mentioned parameters, we also used a mix of \textit{visual} images \textit{combined} with roughness and noise to explore if a combination of the visual and acoustic modalities enabled better identification of the image focus level. This parameter is referred to as \textit{Visual}-plus-\textit{Combined} in the remainder of the paper.

\subsection{Cognitive Workload parameters}
Cognitive Workload is a complex construct of intrinsic, extrinsic and germane load~\cite{chen2016robust}. Hence, we utilized the multidimensional rating design of NASA-TLX~\cite{hart1988development} to acquire explicit cognitive load impressions from users. We employed a subset of the parameters to rate the task difficulty, in particular the three factors of \textit{effort} (Rate the level of effort you needed to put in to complete the task on a scale of 0 (lowest)--4 (highest)), \textit{mental demand} (Rate the level of stress you endured while performing the task on a 0--4 scale)  and \textit{frustration} (Rate the level of frustration you experienced while performing the task on a 0--4 scale). We hypothesized that these three parameters had complementary contributions to the user workload, given the nature of the task.

\subsection{Participants}
20 participants (16 male) with an average age of 28.9 $\pm$ 4.9 years took part in our study. None of them had any formal training in music. Users had to perform the experiment for about one hour, and were financially compensated for their participation. The experimental design was approved by the local ethics committee.

\subsection{Protocol}
The experiment was conducted in two sessions, where each session was divided into six blocks. Each of the six blocks corresponded to one of the visualization parameters (4 acoustic, 1 visual and 1 visual-plus-acoustic). Within each block, the 10 stimuli were played for 2 sec each in random order and repeated thrice, leading to a total of 30 trials (stimulus presentations) in a block.

Session 1 involved playing of a stimulus following which, the user had to immediately rate the corresponding focus level on a scale of 1--10. Session 1 therefore involved \textbf{Immediate Recall} (IR) from the user. In Session 2, the 10 stimuli corresponding to a particular visualization parameter were again repeated thrice in random order, but at the beginning of each of the three repetitions, users were instructed to click a `Yes' radio button if they inferred that the played stimulus corresponds to a particular focus level. The focus level of interest was pre-specified prior to each of the three repetitions, and the objective here was to facilitate perceptual comparisons between the stimuli presented over successive trials to arrive at a decision. Session 2 is similar to the rapid serial visual presentation (RSVP) protocol adopted in psycho-physical studies, and is termed the \textbf{Compared Recall} (CR) session. To enable at least one comparison by the user, we ensured that the target focus level stimulus was not rendered in the first trial following the target focus level specification.   

In the IR session, each stimulus was presented for 2s, preceded by a a fixation cross which was displayed for 500{ms}. The stimulus was followed by a question to let users ascertain the corresponding focus level on a 1--10 scale. A 10s timer was displayed on the screen during the judgement task, and the next stimulus was automatically presented if the user failed to respond within the 10s time-frame. All parameters remained identical in the IR and the CR sessions with the exception of each stimulus being followed by a 3{s} response time in CR (as against a 10s time-frame in IR). 

Upon completion of each block, users were required to rate their experience on the subset of NASA-TLX parameters (mentioned above) to evaluate the cognitive workload. No technical terms were used to represent the various stimuli, and participants were instructed regarding the stimulus type through a coded representation (Type 1--6). Users were also made to perform a practice session before each session, where they were familiarized with all the stimuli employed in the experiments. 

\subsection{Data Acquisition}
As users performed the focus level detection task, we acquired their neural responses (which we hypothesized to capture the cognitive workload experienced) via the 14-channel consumer-grade \textit{Emotiv Epoc} EEG device. Both the IR and CR sessions were split into two halves to calibrate the device regularly, and to prevent participants from experiencing fatigue. The \textit{Epoc} device used has a  128 Hz sampling rate. Our experimental protocol was designed using Matlab PsychToolbox~\cite{brainard1997psychophysics}. 
\section{Explicit Response Analysis}\label{ERA}

\begin{figure*}[h]
\centering
\includegraphics[width=0.49\linewidth]{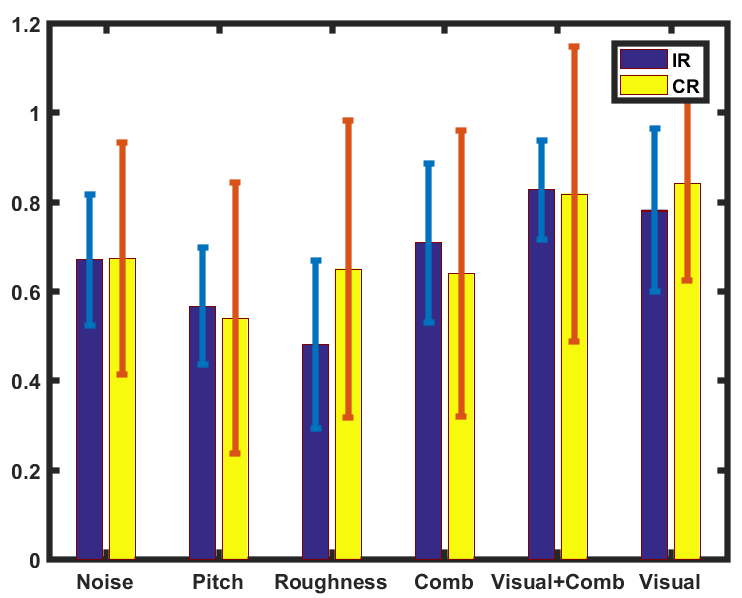}\hspace{0.1cm}\includegraphics[width=0.49\linewidth]{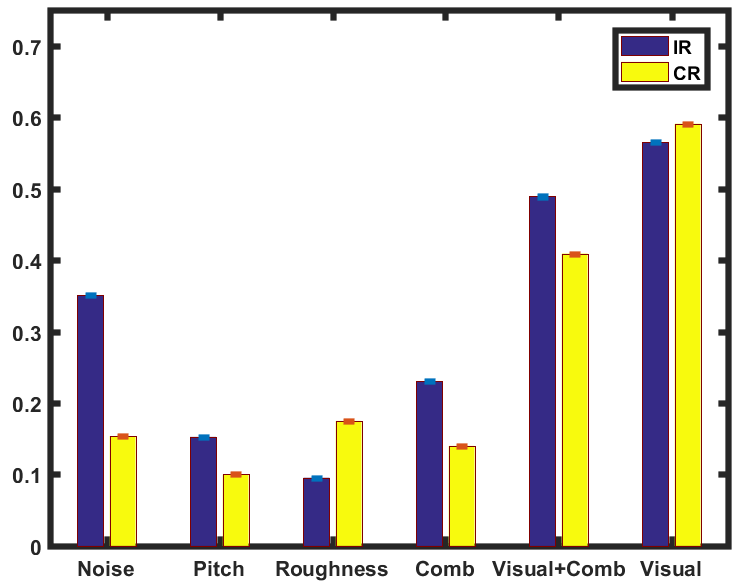}
\caption{(left) Mapping accuracy with the different parameters. Error-bars denote unit standard deviation. (right) Proportion of \textit{low} cognitive load trials for each parameter.}\label{response_plots}\vspace{-.2cm}
\end{figure*}
To begin with, we attempt to model cognitive workload in terms of data:parameter mapping (or recognition) accuracies, and cognitive workload impressions provided by users in terms of the NASA-TLX ratings.  
 
We analyzed the accuracy of user responses for the various stimuli presented and their combinations. We summarize and compare the results with prior work described in~\cite{ferguson2017evaluation}. As with \cite{ferguson2017evaluation}, we considered a~10\% error margin while computing accuracies.

\subsection{Recognition Rate}\label{RR}
For Immediate Recall, the highest mapping accuracy was observed for \textit{visual}-plus-\textit{combined} (0.83 $\pm$ 0.11) while the least accuracy was observed for \textit{roughness} (0.48 $\pm$ 0.19) (see Figure~\ref{response_plots}(left)). This implies that the combination of visual plus acoustic cues facilitated better inference of the image focus level by users. The acoustic stimuli followed a similar trend as in \cite{ferguson2017evaluation} with the \textit{combined} rendering of \textit{roughness} and \textit{noise} (accuracy of 0.71 $\pm$ 0.18) being the easiest to map. The only exception in our study is that \textit{pitch} outperforms \textit{roughness}. The \textit{visual} stimuli (accuracy = 0.78 $\pm$ 0.18) were easier to perceive than all acoustic parameters, in line with one's expectations.

For Compared Recall, very similar mapping accuracies were noted for \textit{visual}-plus-\textit{combined} (0.82 $\pm$ 0.33) and \textit{visual} (0.84 $\pm$ 0.22) (see Figure~\ref{response_plots}(left)). Among acoustic parameters, \textit{Noise} (0.67 $\pm$ 0.26) and \textit{Combined} (0.64 $\pm$ 0.32) were most suitable for detecting the target focus level, while \textit{pitch} was the most difficult (0.54 $\pm$ 0.3). Overall, a similar trend as in \cite{ferguson2017evaluation} was observed. Higher standard deviations in the CR session suggest that focus level identification based on stimulus comparison was more difficult than individually mapping each presented stimulus as in the IR protocol.

\subsection{Cognitive Workload Ratings}\label{cwl_rating}
We considered the mean of the three NASA-TLX parameters (\textit{effort}, \textit{mental demand} and \textit{frustration}) to evaluate the overall mental workload. The average score was thresholded at the mean value of 2 (since the used scale was 0--4) to quantize or characterize a parameter block as inducing \textit{low}/\textit{high} workload. To examine the correlation between recognition rates and the cognitive workload impressions for a given visualization parameter, we computed the proportion of low cognitive workload trials for each of the six parameters. We hypothesized that a visualization parameter that induces low cognitive workload should correspond to higher mapping accuracy, \textit{i.e.}, parameters that correspond to higher accuracies in Figure~\ref{response_plots}(left) will also have many low cognitive load blocks as shown on Figure~\ref{response_plots}(right). To this end, we computed Pearson correlation coeffcients between the mapping accuracies and the proportion of low cognitive load blocks. For IR, we observed a correlation $\rho$ = 0.8339, ${p}$<0.05, while a $\rho$ = 0.8631, ${p}$<0.05 was noted for CR. The above correlations confirm that there exists a congruent relationship between the recognition rates and cognitive workload.

Given that the ultimate objective of this study is to assess mental workload via the implicitly acquired EEG data, we need to label \textit{trials} (denoting the presentation of an individual stimulus) as \textit{high} or \textit{low} cognitive load inducing. To this end, labels were generated for the IR and CR sessions based on user workload impressions. As seen from Figure~\ref{response_plots}(right), we categorized \textit{pitch}, \textit{roughness} and \textit{combined} as \textit{high} load parameters for IR, while \textit{noise}, \textit{visual}, and \textit{visual} plus \textit{combined} as \textit{low} load parameters. This categorization was based on the maximum inter-parameter difference of 0.12 noted between the \textit{combined} and \textit{noise} parameters in Figure~\ref{response_plots}(right). Similarly for CR, the high load parameters included \textit{pitch}, \textit{combined}, \textit{noise} and \textit{roughness} while the low load parameters were \textit{visual}, and \textit{visual} plus \textit{combined}. We observed that the acoustic parameters were rated to induce relatively higher load than the \textit{visual} or \textit{visual} plus \textit{combined} parameters as anticipated.

\begin{figure*}[htbp]
\centering
\includegraphics[width=0.45\linewidth]{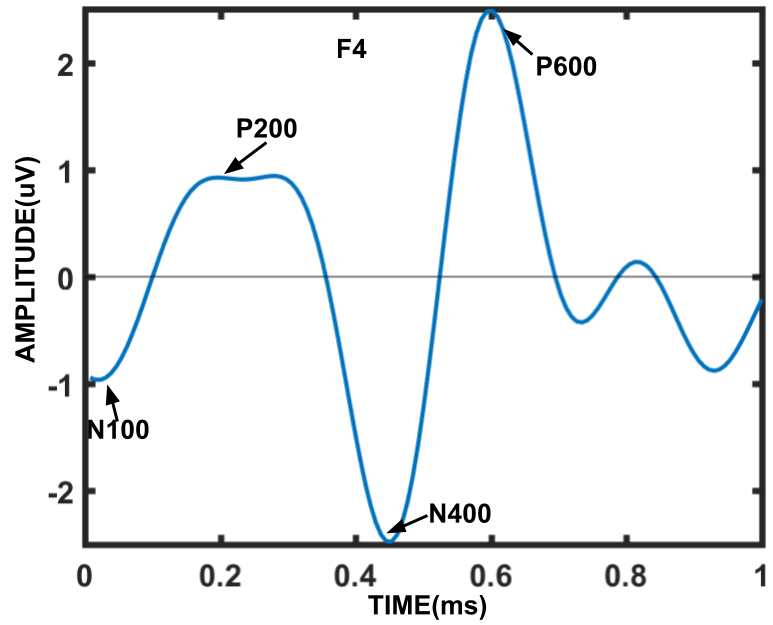}
\hspace{0.5cm}
\includegraphics[width=0.45\linewidth]{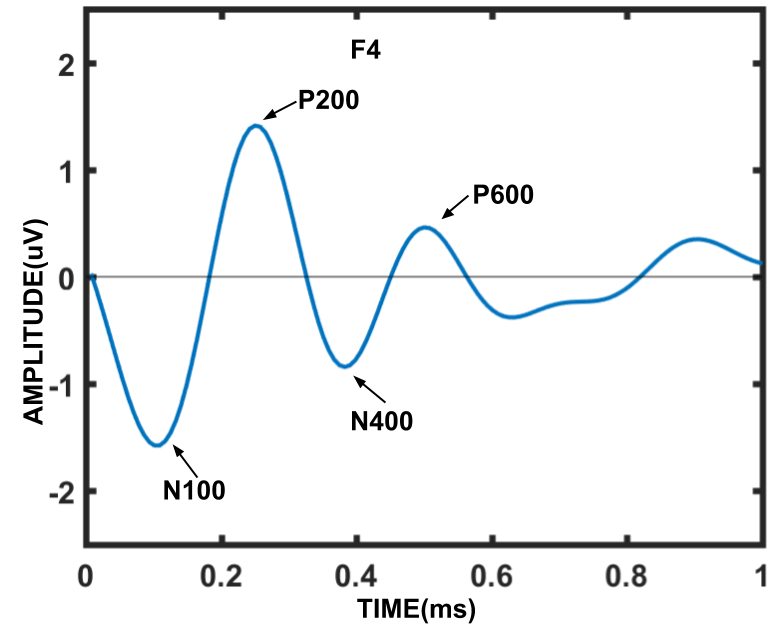}
\vspace{-.3cm}
\caption{ERP plots for \textbf{F4} channel corresponding to \textit{immediate recall} (left) and \textit{compared recall} (right) sessions, with different ERP components labeled. \label{ERPs}}\vspace{.1cm} 
\centering
\includegraphics[width=0.45\linewidth]{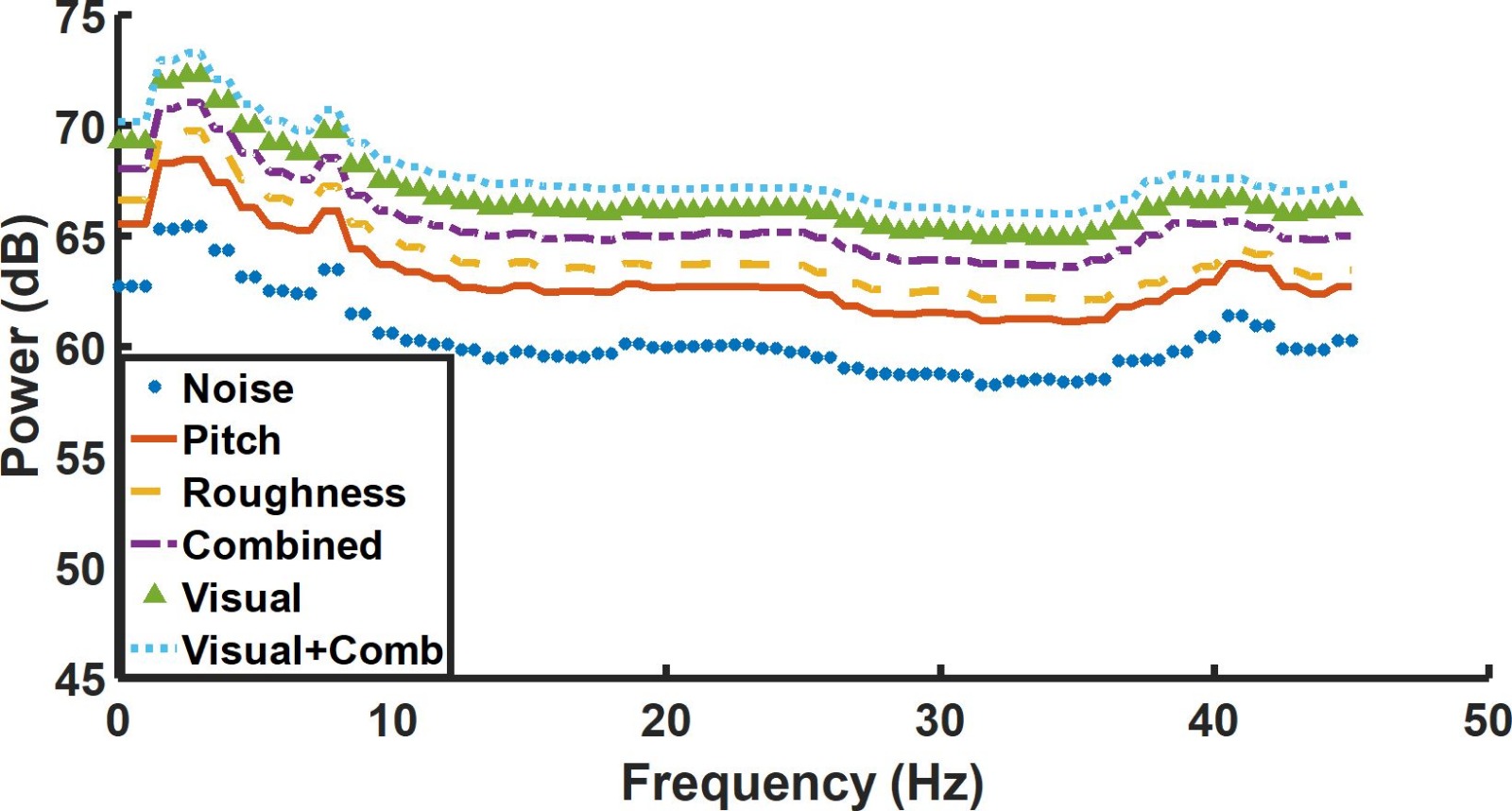}
\includegraphics[width=0.45\linewidth]{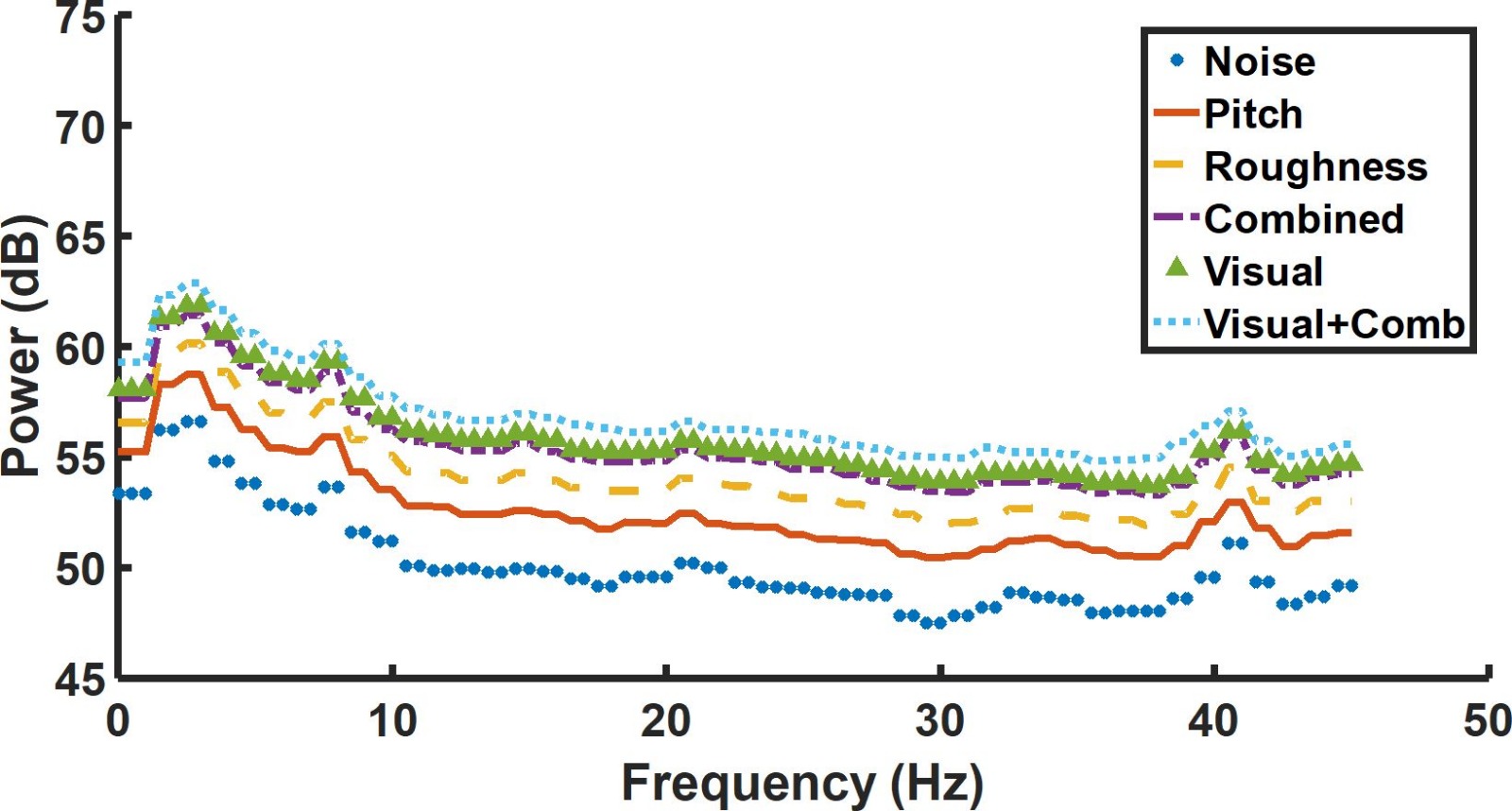}
\caption{Power spectra plots for the different visualization parameters in the IR (left) and CR (right) tasks. \label{pow_spec}}\vspace{-.1in}
\end{figure*}

\section{EEG analysis}\label{EEG}
In this section, we describe the data preprocessing techniques and our inferences from the EEG data acquired during the memory workload evaluation task.

\subsection{Data Preprocessing}
The EEG data acquired using the consumer-grade \textit{Epoc} device is highly susceptible to external electrical noise, motor activity (apart from the task-related activity) such as eye-blinks, head and muscle movements. Thus, the EEG data is subjected to a preprocessing pipeline to eliminate artifacts.

We extracted epochs of 2.5{s} duration for each trial (comprising 0.5{s} of fixation presentation and 2{s} of stimulus presentation). We performed removal of the baseline neural activity DC-offset using the EEG response for the 0.5{s} fixation duration. Further, the EEG signals were subjected to (a) band-limiting between 0.1--45 Hz, (b) visual rejection of noisy epochs and (c) Independent Component Analysis-based rejection of artifacts corresponding to eye-blinks and movements. Muscle  movement  artifacts are removed upon band limiting EEG as they are chiefly concentrated in the  40--100 Hz band, and via manual removal of noisy ICA components.

\subsection{ERP analysis}

Several previous works have investigated Event Related Potentials (ERPs) for human-centric decision tasks like emotion recognition~\cite{bilalpur2017gender}, image annotation~\cite{parekh2018eeg} and error identification \cite{vi2014error}. These markers act as a  bridge between neuroscience and behavioral studies, facilitating better understanding of the structural and functional relevance of neural activity to the task on hand, and validation of the EEG data.

Figure~\ref{ERPs} shows the ERP components observed upon analyzing EEG data from all IR and CR trials. In particular, we emphasize the existence of N100 and P200 components in both the IR and CR sessions. The N100 component is known to reflect processing of acoustic cues, and is typically followed by the P200 component which is fronto-centrally distributed \cite{paulmann2015neurocognition}. Also, amplitude of the N100 component is known to be sensitive to the level of attention~\cite{bellack1998introduction}. We attribute higher N100 amplitude for CR to the experimental design, as the CR protocol forces users to compare the current stimulus against prior ones for target detection, thereby demanding greater attention than the IR protocol (as also reflected via higher variance in mapping accuracies for CR). In addition to the N100 and P200 components, we also encountered the N400 and P600 components in the F4 channel. Several works~\cite{kutas2011thirty,brouwer2017proper} confirm the presence of N400, P600 components in language comprehension and semantic memory understanding tasks. However, their presence while processing visualizations has not been explored as yet.

\subsubsection{Spectral Analysis}
Prior studies on memory workload for the $n$-back task~\cite{wang2016using} have demonstrated the existence of differences in $\alpha$~(8-13 Hz), $\theta$ (4-7 Hz), lower $\beta$ (13-16 Hz) and higher $\gamma$ (40-45 Hz) EEG band powers with varying task difficulty. We investigated the power spectrum averaged over all the EEG channels for any existence of such cues. The power spectrum analysis (Figure~\ref{pow_spec}) suggests an increased activity in the $\alpha$ band (as well as other frequency bands such as $\delta$ (1-4 Hz) and $\gamma$) for low workload parameters identified from Section~\ref{cwl_rating}. {These patterns are better observable for the CR task as compared to IR.} Differences noted in~\cite{wang2016using} for the $\theta$ band are not observed in our study. It is also interesting to note that the power spectral trends for both the IR and CR tasks are similar indicating that memory workload is reliably captured by our study. Some similarities can be noted between Figure~\ref{response_plots}(right) and Figure~\ref{pow_spec}. Those parameters associated with \textit{low} cognitive workload (such as \textit{Visual} and \textit{Visual} plus \textit{Combined}) are associated with higher spectral power (in particular the $\alpha$ band) than \textit{high} workload parameters (such as \textit{pitch} and \textit{noise}).
\section{Cognitive Load Classification}\label{CLC}
This section describes how we train classifiers from the compiled EEG data and cognitive workload labels to implicitly assess memory workload from EEG. To this end, we labeled all trials corresponding to parameters associated with \textit{low}/\textit{high} workload (from Section~\ref{cwl_rating}) for the IR and CR tasks accordingly, \ie, \textit{Visual}-plus-\textit{Combined}, \textit{Visual} and \textit{Noise} parameters were considered to be low workload inducing for IR, while only \textit{Visual} and \textit{Visual}-plus-\textit{Combined} were deemed as low workload inducing for CR.


\subsection{Classification methods}
\subsubsection{Traditional methods}

\begin{figure*}
\vspace{-0.5cm}
\centering
\includegraphics[width = 0.95\linewidth]{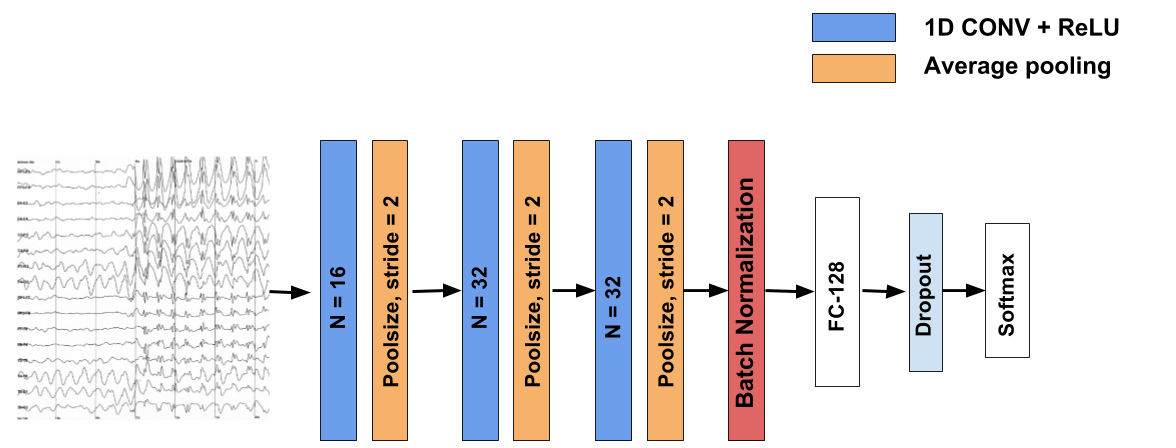}
\caption{CNN architecture showing various layers in the model and parameters.}
\label{DLarch}
\end{figure*}

\begin{sloppypar}
We employed traditional machine learning methods-- Naive Bayes (NB), Linear Discriminant Analysis (LDA), SVM with linear kernel (LSVM), and SVM with RBF kernel (RSVM) for EEG-based \textit{low}/\textit{high} workload categorization. Prior to classification, the 14 channel, 2s long stimulus-duration data was vectorized and subjected to dimensionality rejection via principal component analysis to retain 90\% variance. We performed 10 repetitions of 5 fold cross validation on the data. Given the class imbalance for the IR and CR tasks, we chose F1-score as the metric to evaluate classifier performance.
\end{sloppypar}

\subsubsection{Convolutional Neural Network} 
The CNN feature learning can achieve better dimensionality reduction by projecting the EEG data onto a robust feature space for preserving task relevance and invariance to noise.

We adopted a 3-layer CNN model proposed in~\cite{rad2018deep} to learn a robust representation from EEG data for classification of cognitive workload. Three convolution together with rectified linear unit (ReLU) activation function and average pooling layers are stacked to extract task specific features (Figure~\ref{DLarch}). The convolutions employed are 1-dimensional along time. Batch normalization~\cite{ioffe2015batch} is used after the third CNN layer to reduce the internal covariate shift and accelerate the training. To prevent over-fitting, we used dropout after the fully connected layer with 128 neurons. We finally classify with \textit{softmax} over 2 output neurons. The number of kernels increase with the depth of the convolution network as analogous to the VGG architecture~\cite{simonyan2014very}. We have optimized the network for categorical cross entropy using vanilla stochastic gradient descent with Nestrov momentum and weight decay. CNN hyper-parameters are specified in Table~\ref{network_hyp}. The values for these hyper-parameters are mainly adopted from~\cite{rad2018deep} or otherwise decided by cross-validation (10 repetitions of 5-fold). 

\begin{table}[]
\fontsize{7}{7}\selectfont
\centering
\caption{Convolutional neural network parameters.} \vspace{-.2cm}
\scalebox{1.0}{%
\label{network_hyp}
\begin{tabular}{ll}
\hline
\hline
\textbf{Parameter }       & \textbf{Value }   \\ \hline
Learning rate          & 0.01     \\
Kernel size            & 3        \\
Stride size            & 2        \\
Pool size              & 2        \\
Batch size             & 32       \\
\# Kernels(layer wise) & 16,32,32 \\
Momentum			   & 0.9	  \\
Weight decay		   & 0.0001   \\
Dropout                & 0.1     \\ 
\hline \vspace{-.5cm}
\end{tabular}}
\end{table}

\begin{table*}[h]
\centering
\fontsize{8}{8}\selectfont
\caption{Classification results summary (* denotes F1-score distribution significantly above chance level with $p$<0.05).}
\label{results} \vspace{-.2cm}
\scalebox{1.3}{
\begin{tabular}{|l|lll|lll|}
\hline

 & \multicolumn{3}{c|}{\textbf{IR}}       & \multicolumn{3}{c|}{\textbf{CR}}       \\ \hline
\textbf{Classifier}& \multicolumn{1}{c}{\textbf{Precision}} & \multicolumn{1}{c}{\textbf{Recall}} & \multicolumn{1}{c|}{\textbf{F1-score}} & \multicolumn{1}{c}{\textbf{Precision}} & \multicolumn{1}{c}{\textbf{\textbf{Recall}}} & \multicolumn{1}{c|}{\textbf{F1-score}} \\ \hline
\textbf{NB}         &    0.48 $\pm$ 0.01       &   0.48 $\pm$ 0.03     &    0.48 $\pm$ 0.02      &    0.49 $\pm$ 0.06       &    0.37 $\pm$ 0.06    &     0.42 $\pm$ 0.04     \\ 
\textbf{LDA}        & 0.50 $\pm$ 0.01	& 0.61 $\pm$ 0.02	& 0.55 $\pm$ 0.02\text{*} &	0.56 $\pm$ 0.05	& 0.50 $\pm$ 0.07	& 0.53 $\pm$ 0.07\text{*}\\ 
\textbf{LSVM}       & 0.48 $\pm$ 0.01	& 0.54 $\pm$ 0.02	& 0.51 $\pm$ 0.02\text{*} & 0.58 $\pm$ 0.04 &	0.52 $\pm$ 0.07	& \textbf{0.55 $\pm$ 0.06}\text{*}\\ 
\textbf{RSVM}       & 0.37 $\pm$ 0.04	& 0.46 $\pm$ 0.15	& 0.41 $\pm$  0.03	& 0.34 $\pm$ 0.01	& 0.55 $\pm$ 0.00 &	0.42 $\pm$ 0.00\\ 
\textbf{CNN}        &0.64 $\pm$ 0.02	& 0.64 $\pm$ 0.02	& \textbf{0.64 $\pm$ 0.02}\text{*}	& 0.54 $\pm$ 0.10	& 0.52 $\pm$ 0.06	& 0.52 $\pm$ 0.07\text{*}\\ \hline

\end{tabular}}
\vspace{-.1cm}
\end{table*}

\subsection{Results}
{Table~\ref{results} summarizes the performance of various classifiers in terms of F1-scores ($\mu \pm \sigma$) obtained over 50 runs. Precision and recall values are also tabulated for better insights. The obtained results clearly show that the LDA, LSVM and CNN classifiers outperform Naive Bayes and RSVM, and achieve better-than-chance cognitive load classification. The fact that LSVM outperforms RSVM for both IR and CR tasks implies that non-linear kernels do not result in better classification performance for our data. Traditional classifiers result in low precision for the IR task, and low recall for the CR task. Overall, the CNN classifier achieves the most balanced performance in terms of precision and recall, and produces the best workload classification performance for the IR task (F1 = 0.64). Nevertheless, its performance decreases for the IR task presumably because of the class imbalance and the fact that training data was far fewer for this condition (4 \textit{low} cognitive load vs 2 \textit{high} cognitive load parameters, and only those trials eliciting a user response were employed for training). Correspondingly, higher variance in the performance of all classifiers is noted in the CR condition. The LSVM classifier performs best for the CR task. Overall, our classification results emphasize the need for efficient and robust feature learning for workload estimation from noisy EEG signals.}

\section{Discussion and Conclusion}\label{DC}
This work examines the efficacy of the \textit{pitch}, \textit{roughness}, \textit{noise}, and \textit{combined} acoustic (in combination with visual) parameters for conveying image focus level in terms of explicit data:parameter mapping accuracies and user cognitive load impressions conveyed via NASA-TLX attributes. Specifically, parameters associated with \textit{low} cognitive load result in higher mapping accuracies in line with one's expectations. Furthermore, we perform automated classification of cognitive load from EEG signals acquired via a commercial wireless device, and labels based on explicit responses to achieve a maximum (and significantly above-chance) F1-score of 0.64. In terms of novelty, our work improves over evaluation studies conducted in~\cite{ferguson2017evaluation,ferguson2018investigating} which exclusively rely on explicit user responses for workload understanding.


On the whole, our findings mirror those in~\cite{ferguson2017evaluation} and~\cite{ferguson2018investigating}. The best performing acoustic parameter in terms of data:sound mapping accuracies is the \textit{combined} rendering of \textit{roughness} and \textit{noise} (Section~\ref{RR}), which owing to its negative attributes best conveys \textit{image blur} (inverse of \textit{focus level}). We observe that focus recognition accuracy is higher for {visual} stimuli than acoustic stimuli; nevertheless, the recognition accuracy is highest for \textit{visual}-plus-\textit{combined} implying that acoustic information augments visual information towards determining the focus level. Congruence between recognition accuracies and cognition load is also highly noticeable. Specifically, the \textit{visual} and \textit{visual}-plus-\textit{combined} conditions are reported to induce \textit{low} memory workload over a majority of the trials, while \textit{noise} is observed to be the most intuitive in terms of acoustic parameters based on NASA-TLX responses. 

We designed two different tasks for determining the image focus level in our experimental design-- immediate recall, where users had to immediately detect focus level from the presented stimulus, and compared recall where the user needed to detect a pre-specified target level from among serially presented stimuli. We hypothesized that CR would be more facile for visualization understanding since comparisons would enable a better assessment of the rendered visualizations. However, different from our expectation, a greater variance in IR recognition accuracies suggests that focus level identification was perhaps more difficult for the IR task. It is also pertinent to note that the N100 ERP component was stronger for the IR task, and indicative of the greater cognitive attention required in this setting. A more thorough analysis of the focus levels that were easy/difficult to detect in the IR and CR settings is required as part of future work.

Power spectral analysis also confirmed that low workload parameters are characterized by higher $\alpha$, $\delta$ and $\gamma$ band powers as compared to high workload parameters, and that the power spectral densities for the CR task are significantly lower than those for the CR condition. In terms of binary cognitive workload classification, the CNN model produces the most balanced performance in terms of precision and recall, and results in the peak F1 score of 0.64 for the IR condition. Relatively lower F1 scores achieved for CR can be attributed to fewer training data available in this condition (about $\frac{1}{7}^{th}$ of the training data available for IR owing to the experimental design). While this work compiles data from the general population, data sonification could specifically benefit the visually impaired community, and future work would evaluate sonification techniques on such participants.
\section*{Acknowledgements}
We thank Steve Brewster and Jamie Ferguson for kindly providing us with the visual and auditory data used in their experiments. This research is supported by the National Research Foundation, Prime Ministers Office, Singapore under its
International Research Centre in Singapore Funding Initiative. Maneesh Bilalpur completed this work as a visiting student at the SeSaMe Centre in the National University of Singapore.

\bibliographystyle{ACM-Reference-Format}
\balance
\bibliography{sample-bibliography}

\end{document}